 \documentclass[doublecol]{epl2} 

\title{Wave-induced  motion  of  magnetic spheres}

\author{Christophe Gissinger\inst{1}}

\institute{                    
  \inst{1} Laboratoire de Physique Statistique, Ecole Normale Supérieure, CNRS, Université P. et M. Curie, Université Paris Diderot - Paris, France}
\pacs{05.45.-a}{Nonlinear dynamics and chaos}
\pacs{45.40.-f }{Dynamics and kinematics of rigid bodies}
\pacs{47.63.-b}{Biological fluid dynamics }
\pacs{85.70.Rp}{Magnetic levitation, propulsion and control devices}

\abstract{We report an experimental study of the motion of
magnetized beads driven by a travelling wave magnetic
field. For sufficiently large wave speed, we report the existence of a
backward motion, in which the sphere can move in the direction opposite
to the driving wave. We show that the transition to this
new state is strongly subcritical and can lead to chaotic motion of
the bead. For some parameters, this counter-propagation of the sphere
can be one order of magnitude faster than the driving wave
speed. These results are understood in the framework of a 
model based on the interplay among solid friction, air resistance and magnetic
torque.}

\begin{document}

\maketitle

\section{Introduction}

Many situations, in laboratory or in Nature, involve the locomotion of
 spherical particles on a solid plane through a viscous fluid \cite{Michaelides97}. Despite
the apparent simplicity of the problem, there are still many
unsolved questions concerning the exact mechanisms controlling the
motion in these systems, such as the transition from static to rolling
friction, the role of surface elasticity, or the hydrodynamical interaction between the fluid  and the moving object.

During the last decade, this type of problem has gained some renewed
interest in the field of microfluidics. For instance,  biomedical applications involve microrobots such as artificial micro-swimmers made of super-paramagnetic beads subject to an external magnetic field \cite{Peyer2013,Pamme2006}. 
In some cases (so-called 'surface walkers'), it has been reported that actuation of the microrobot can be controlled by the presence of a surface wall \cite{Sing2010,Karle2011}.
On the other hand, the problem of particles transported by electromagnetic waves is widely studied in  the framework of the plasma-electrons interaction. Surprisingly, only few studies have been done on electromagnetic control of
particles on macroscopic scales \cite{Bolcato93}.
In this letter, we focus on the motion of macroscopic magnetic beads on a flat horizontal surface driven by the magnetic
force due to a travelling wave. \\

\begin{figure}
\centerline{\includegraphics[width= 7cm]{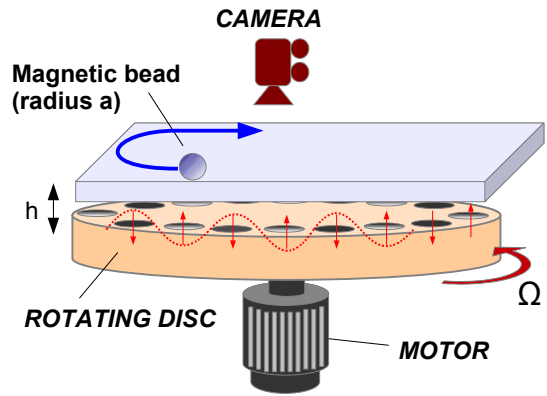} }
\caption{Experimental set-up. A  bead moves on a flat surface under the effect of
    a traveling magnetic wave created by $16$ Neodymium magnets located on a rotating disc.}
\label{schema}
\end{figure}

\section{Experiment}

Fig.\ref{schema}-left shows a schematic picture of the experiment: a
spherical bead is placed on a fixed horizontal plane made of methyl-methacrylate (plexiglass)
of width $a_p=5$mm. The bead, of radius $a$, is made either of
weakly ferromagnetic mild steel  or Neodymium NdFeB (permanent magnet) depending on the experiments and can freely move on the surface. At a distance $h$ below the plane is located a rotating disc
containing $16$ Neodymium magnets disposed with a regular spacing
along a circle of radius $R=83$mm. These magnets are cylinders of
diameter $d_m=20$mm and height $h_m=10$mm, generating a magnetic field
of $B_m^0=0.45T$ at their surface. The magnets are arranged such that two
adjacent magnets, separated by a distance $d_m=2\pi R/16=32.5$mm, are
oriented with opposite polarity. This rotating disc therefore
generates a sinusoidal magnetic field propagating in the azimutal direction with a pulsation
$\omega=2\pi f_d/16$ and a wavenumber $k=\pi/d_m$, where $f_d$ is the
rotation rate of the disc.
 The bead therefore moves with a velocity $V$ (normalized by the wave speed $c$) under the influence of this travelling magnetic field along circular trajectories which are
recorded by a fast camera located on the top of the set-up. By changing the distance $h$ between the rotating disc and
the fixed plate, one can also experimentally vary the magnitude of the magnetic field applied to the sphere.

\begin{figure}
\centerline{\includegraphics[width= 9 cm]{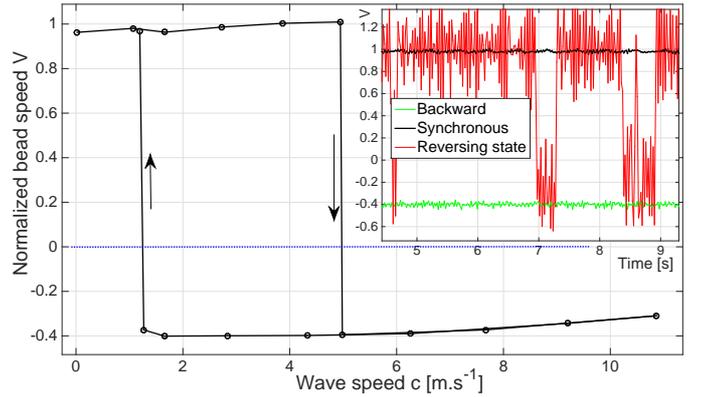} }
\caption{Bifurcation diagram of the velocity $V$ of a mild steel bead
  of radius $a=3$mm and disc/plate distance $h=4$mm. Note the
  coexistence of two solutions, corresponding to either synchronism
  with the wave or counter-propagating motion. Typical time series
  are shown in inset (see text). }
\label{hysteresis}
\end{figure}

Figure \ref{hysteresis} shows the evolution of the normalized
velocity $V$ of a mild steel bead of radius $a=3$mm as a function of
the  wave speed $c$. At small disc frequency, the bead moves in
synchronism with the wave ($V=1$, upper branch). This synchronous translation is relatively intuitive: under the effect of the applied field, the magnetized bead is trapped inside the potential well of one of the
magnet of the disc. As the disc rotates, the  sphere therefore slides above the magnet without
rolling and keeps its magnetic moment aligned with the local field (time series in the inset figure, black curve, for $c=3$m.s$^{-1}$). 

However, for larger wave speed ($c>5$m.s$ ^{-1}$), the system suddenly
bifurcates to a very different solution, in which the bead now 
propagates in the {\it backward} direction related to the driving
wave  (typical time series in green, for $c=6.3$m.s$^{-1}$).
 Note that this backward motion is in this case slower than the wave speed ($V\sim
-0.4$), and  slightly decreases with $c$.

If the wave speed is decreased from this new state, the transition back to synchronous positive translation is obtained for much smaller velocity, thus showing that this bifurcation is
subcritical and associated with a strong hysteresis.

For some parameters, the bistability between these two solutions can
yield complex non-linear behavior. The red curve in the inset of Fig.\ref{hysteresis} 
shows a time series obtained for $h=8$mm, $a=2.5$mm, $c=0.68$m.s$^{-1}$. For these
parameters, both positive and negative $V$ solutions are accessible and
the system undergoes chaotic switches between the two branches.\\

\begin{figure}
\centerline{\includegraphics[width= 9 cm]{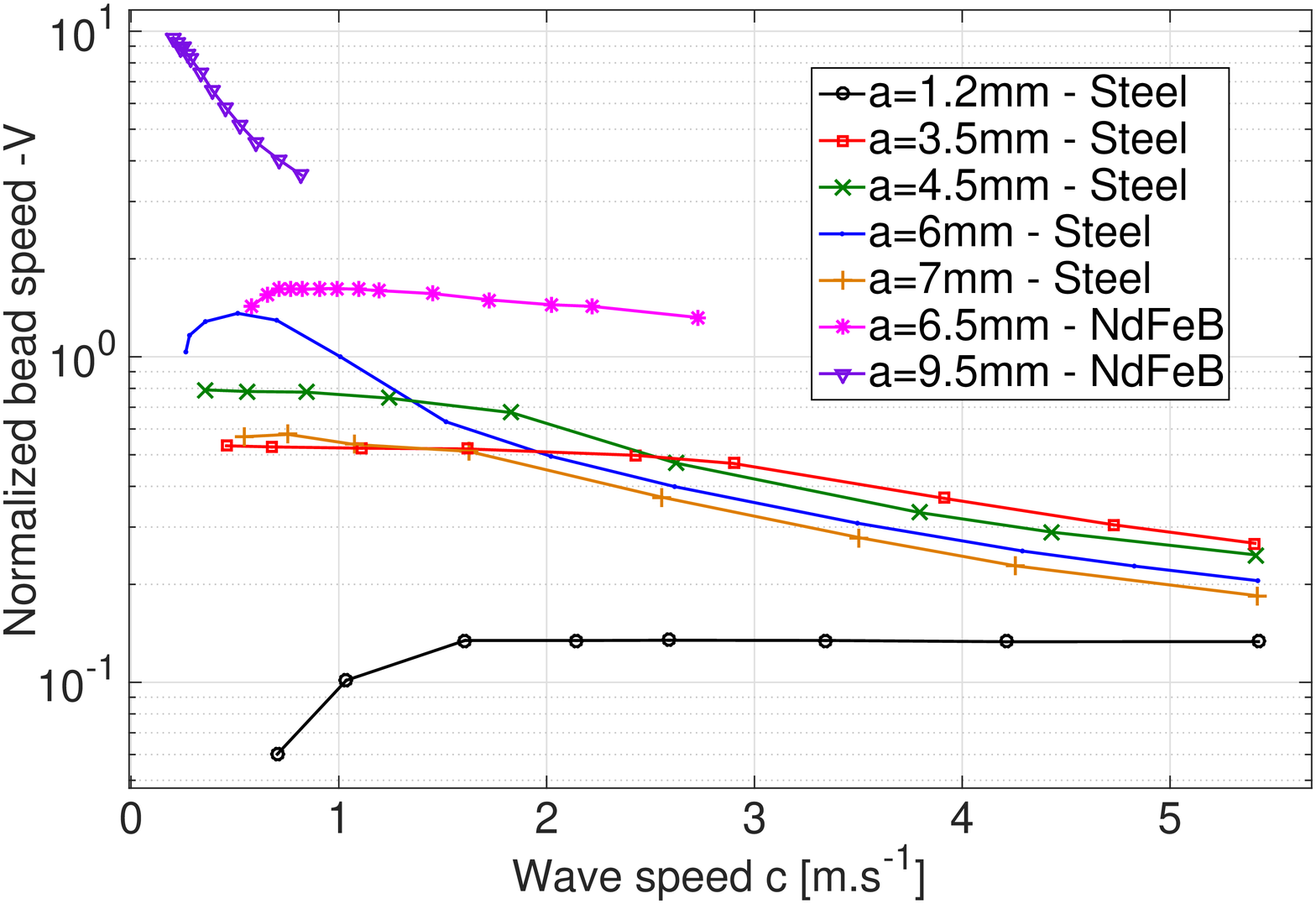} }
\caption{Velocity of counter-propagative beads as a function
  of the wave speed $c$, for $h=10$mm. For some parameters, the backward velocity can be $10$
  times higher than $c$.}
\label{bifs}
\end{figure}

In fact, the evolution of the counter-propagative solution strongly
depends on the parameters used in the experiment. Fig.\ref{bifs} shows
the evolution of $-V$ (only the negative solution is shown) as a function of $c$ for $h=10$mm and for different values of the sphere radius. For the smallest value of $a$, the velocity $V$ is constant,
whereas it systematically decreases with $c$ for larger $a$. Note that the
maximum velocity reached by the sphere first increases with $a$,
before decreasing for the largest values of $a$. For $a=6$mm, this
maximum velocity reaches $V=-1.2$ (for $c=0.5$m.s$^{-1}$), meaning
that the bead is travelling backward {\it faster} than the magnetic
wave driving its motion.

If mild steel beads are replaced by spherical Neodymium $NdFeB$ magnets,  the
magnetic moment of the sphere is now fixed, rather than induced
by the applied field. Such magnetic spheres show the same type of
behavior discussed above, with a bistability between forward and
backward solutions. Note however in Fig.\ref{bifs} that
the maximum negative velocity can be very large: for $a=9.5$mm for
instance, the sphere propagates backward $10$ times faster than the
forward magnetic wave associated to the rotating (See Supplemental Material at URL for videos)\\



\begin{figure}      
                  \includegraphics[width= 8 cm]{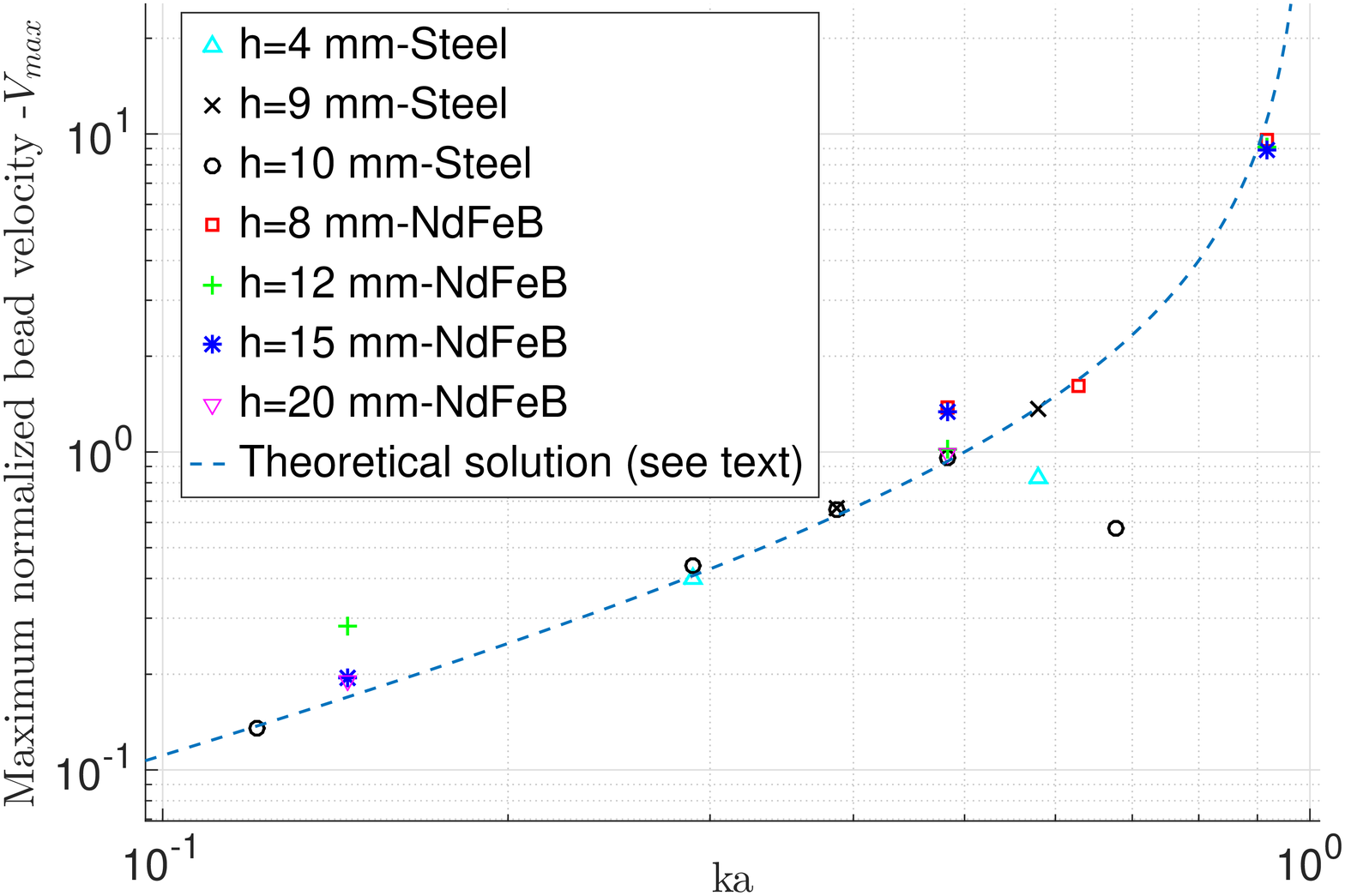}
            \label{bif_ka}
             \vskip\baselineskip 
         \includegraphics[width= 2.5 cm]{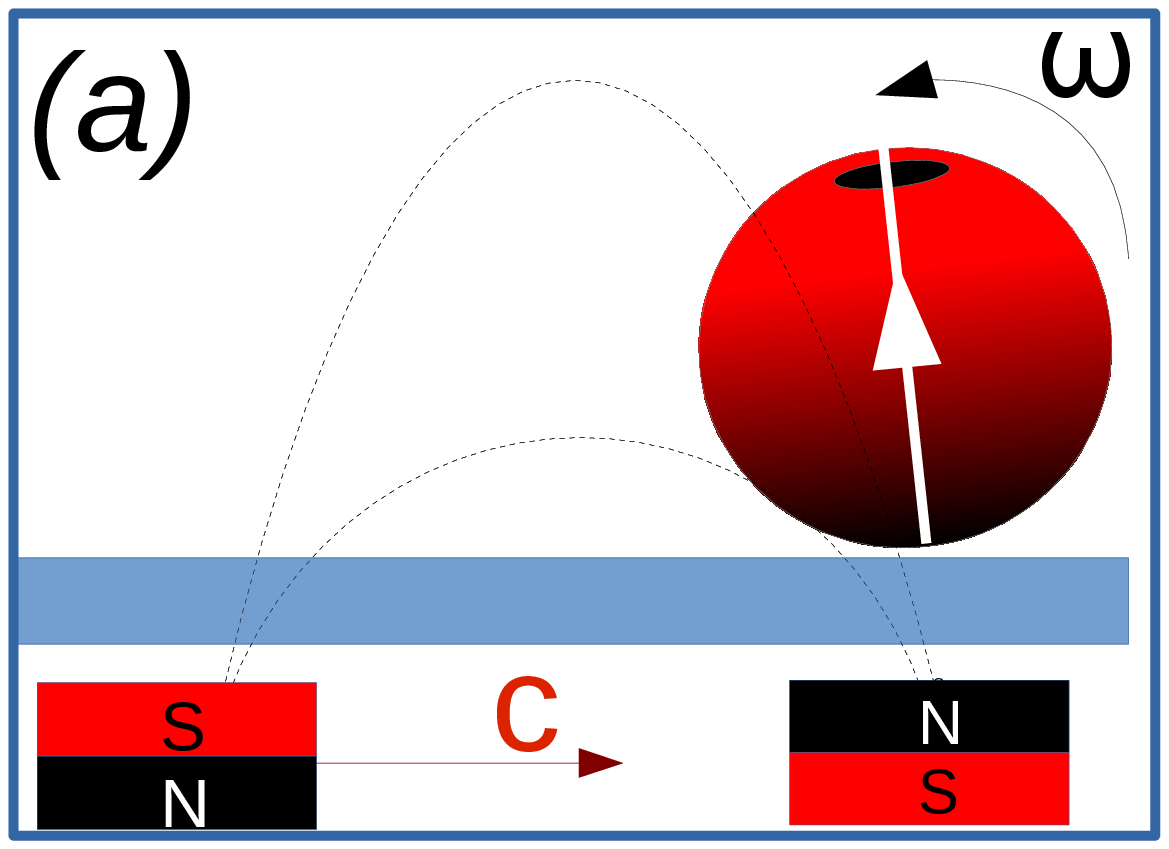}
             \label{schema_a}
               \hspace{3 mm}
          \includegraphics[width= 2.5 cm]{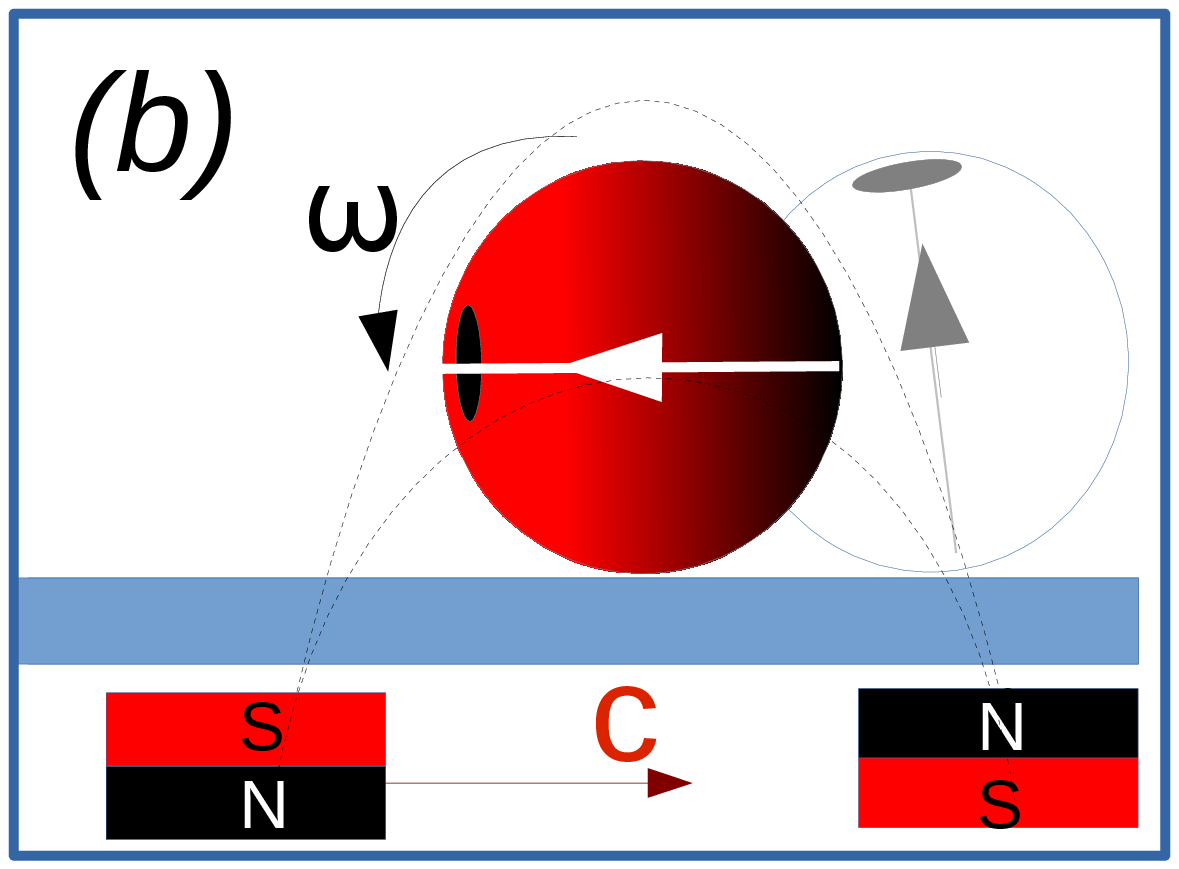}
            \label{schema_b}
                \hspace{3 mm}
      \includegraphics[width= 2.5 cm]{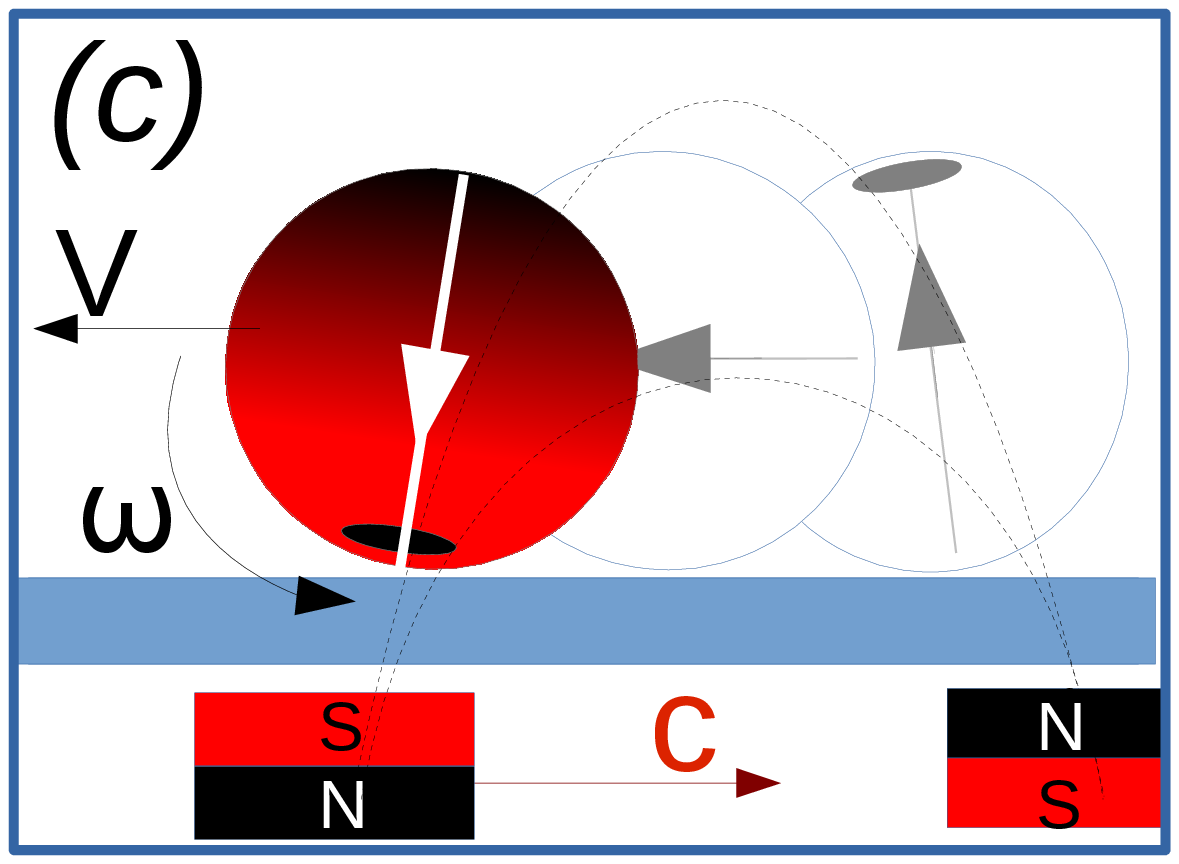}
            \label{schema_c}          
        \caption{Top: maximum velocity of the counter-propagative solution as a
  function of $ka$. Dashed line corresponds to solution (\ref{solutionA}). Bottom:
  simple picture of the mechanism leading to this solution. As the bottom magnets move to the right at speed $c$, the sphere rotates counter-clockwise in order to stay aligned with the field, producing a translation to the left at a speed $v>c$ .}
        \label{schema_ka}
    \end{figure}

\section{Theoretical model}

In order to clarify  the mechanism by which such a solution occurs, a simple one-dimensional model is now derived, in which the sphere now moves along some horizontal $x$ coordinate at a distance $z$ from the bottom magnetic wave.
Let us suppose that the magnetic field due to the rotating disc (black lines in Fig.\ref{schema_ka}-bottom) can be
modeled by a magnetic wave ${\bf B}=B_0e^{-kz}\left[\cos{(\omega_0t-kX)}{\bf e_x}+\sin{(\omega_0t-kX)}{\bf e_z}\right]$, where $X$ represents the local position along the 
trajectory, $k$ is the wavenumber of the magnetic wave and $z$ is the distance from the disc. The bead of mass $M$ is also supposed to be uniformly magnetized
(Magnetization density $M_0$) with a dipole moment ${\bf m}=
  m_0\left[\cos\theta(t){\bf e_x}+\sin\theta(t){\bf e_z}\right]$ of constant amplitude $m_0$ but with a direction
which depends on the orientation of the sphere (white arrow in Fig.\ref{schema_ka}-bottom). Both ${\bf B}$ and ${\bf
  m}$ are supposed to lie in the $(x,z)$ plane. The
horizontal component $F_x$ of the magnetic force ${\bf F}={(\bf m.\nabla}){\bf B}$ and the magnetic torque $ \Gamma_B={\bf m \times B}$ acting on the particle are given by:

\begin{equation}
F_x=km_0B_0\sin\left( \varphi-\theta \right) \hspace{1 cm}
\Gamma_B=m_0B_0\sin\left( \varphi-\theta \right)
\end{equation}

where $\varphi(t)=\omega_0t-kX(t)$ is the phase of the magnetic field
at point $X$ and time $t$. During its motion, the sphere is also
subject to the solid friction $r_x$ from the surface on which it is
moving and to the fluid friction from the air $F_a=\frac{1}{2}C_D\pi
a^2\rho_{air}U^2$sign$(U)$, where $U={\dot X}$ is the velocity of the
bead and $C_D$ is the drag coefficient for a sphere rolling on a plane. The motion of the sphere is then governed by Newton's law for
both the velocity $V$ of the center of mass  and the
angular velocity $\omega$ with respect to the same center of mass. By using the wave speed $c=\omega_0/k$ as a typical velocity scale and $l_0=a$ as a typical length scale, the equations of motion reduced to the simple dimensionless system :

\begin{eqnarray}
{\dot V}&=&ka\tau\sin\psi +R -KV^2sign(V)\label{eq1}\\
\frac{5}{2}{\dot \omega}&=&\tau\sin\psi +R \label{eq2}\\
{\dot \psi}&=&ka-kaV-\omega \label{eq3}
\end{eqnarray}

where $R=r_xa/Mc^2$ is the dimensionless (unknown) solid friction, $\psi=\varphi-\theta$ is the difference of phase between the
magnetic field at point $X$ and the magnetic moment of the sphere.  $K$ is given by $K=\frac{3}{8}C_D\rho_{f}/\rho$, where $\rho_f$ and $\rho$ are respectively the density of the surrounding fluid and of the bead. For simplicity, the drag torque has been neglected in eq.(\ref{eq2}), since it is expected to be much smaller than the translational drag. In addition, we assume no deformation of the plane during the bead displacement, so rolling friction does no work. 
Three dimensionless numbers therefore control the problem: $ka$, $K$ and $\tau=M_0B_0/\rho c^2$, which compares the magnitude of the magnetic driving to inertial effects. 

We now search for for stationary solutions,
$\dot{\psi}={\dot \omega}=\dot{V}=0$. First, we are interested in the
solution for which the sphere rolls without sliding, corresponding to
$\omega=-V$ in our dimensionless variables. Eq.(\ref{eq3}) and
combination of eqs. (\ref{eq1}) and (\ref{eq2}) leads to the solution:

\begin{equation}
V_0=-\frac{ka}{(1-ka)} ,\hspace{2 mm} \sin\psi_0=\frac{Kka^2}{\tau(1-ka)^3},  \hspace{2 mm} R_0=\frac{-K ka^2}{(1-ka)^3} 
\label{solutionA}
\end{equation}

In Fig.\ref{schema_ka}-top,  this theoretical backward solution $V_0$ is plotted as a function of $ka$ (dashed line)  and compared to the maximum amplitude of the experimental backward solution  obtained for different values of $h$ (i.e. different values of the magnetic field magnitude). It appears that most of the points  collapse on the theoretical prediction $V_0$, independently
of the values of $a$, $h$ and the bead materials.

Solution ($\ref{solutionA}$)  has a simple physical meaning: In
the 'local' center of mass reference frame, the sphere experiences a magnetic
field rotating at a pulsation $\dot{\varphi}=\omega_0-kU$, i.e. the
local rotation frequency of the field is doppler-shifted by the displacement of
the bead. Under the effect of this (locally) rotating  field, the
magnetic moment of the sphere locks to the field and a synchronous
rotation $\dot{\theta}=\dot{\varphi}$ is achieved. Since the sphere
rolls without sliding, this positive angular rotation enhances the
translational counter-motion of the sphere, increasing in return the
doppler-shift of the wave speed (see pictures of Fig.\ref{schema_ka}). 

An important point is that this motion is not bounded by the velocity of the travelling
field: as $ka$ tends to $1$ (i.e. when the half-perimeter of the bead equals the distance between two adjacent magnetic poles of the wave), only an infinitesimal displacement of the magnetic wave is necessary for the sphere to travel from one pole to the other while keeping constant its angle with the local magnetic field. In other words, the sphere propagates along the field lines several times faster than $c$ in order to conserve synchronous rotation with the local magnetic field.\\

Note that the above argument is only valid for a sphere rolling without sliding, the corresponding rolling friction $R_0$ being smaller than the sliding friction. For a sphere which slides  during its motion, the friction $R$ on the bottom surface is rather given by $R_S=-\mu W$, where $\mu$ is the  friction coefficient and $W=ka\tau\cos{\psi}$ is the dimensionless normal force due to the vertical magnetic attraction force (gravity being neglected).
 
 A transition from pure rolling to sliding is expected when the rolling resistance $R_0$ exceeds the sliding friction $R_S$. By using the expression (\ref{solutionA}) for $R_0$, this leads to the definition of a new dimensionless number controlling the motion of the bead:

\begin{equation}
N=\frac{|R_0|}{|R_S|}=\frac{K\tau}{\mu}\frac{ka}{(1-ka)^3}
\label{N}
\end{equation}

where the limit $\psi\ll1$ (small phase lag) has been taken.
When $N>1$, the fast solution $V=V_0$ disappears and the sphere starts sliding during its backward motion. By injecting the sliding friction $R_S$ into equations (\ref{eq1}-\ref{eq3}), one find the following solution:

\begin{equation}
V_S=\frac{V_0}{\sqrt{N}} ,\hspace{5 mm} \sin\psi_S=\mu ka
\label{solutionB}
\end{equation}

This second solution still corresponds to a bead propagating backward and a synchronous angular rotation, but due to some sliding, the sphere now moves slower than the pure rolling solution $V_0$.

\begin{figure}
\centerline{\includegraphics[width= 9 cm]{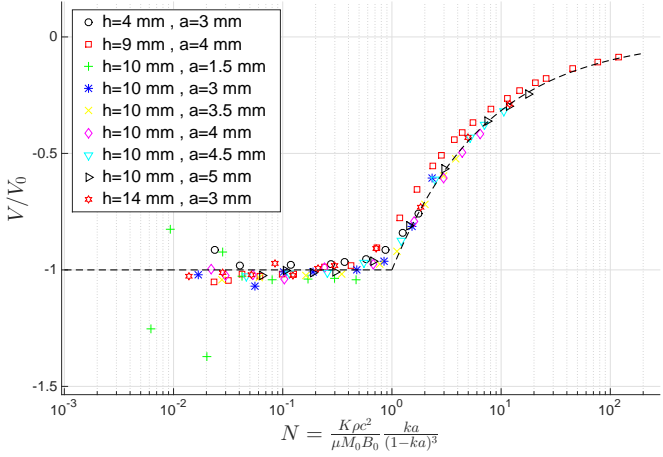}}
\caption{Evolution of $V/V_0$ as a function of $N$ for mild steel beads. Most experimental  data collapse on the dashed line corresponding to model
  equations (\ref{solutionA}) and (\ref{solutionB}), i.e. pure rolling for $N<1$, and rolling-sliding for $N>1$. }
\label{rescaling}
\end{figure}

\section{Comparison between theory and experiment}

At this point it is interesting to compare the model derived above
with our experimental setup.  When the mild steel bead  is first put on the table, it is
magnetized by the magnetic field below, such that its
magnetization density is proportional to the ambient magnetic field.
With typical values of the experiment ($B_0\sim 300$G and $c \sim 3$m.s$^{-1}$), $\tau$ ranges between $10^{-2}$ and $100$. 
Some experiments \cite{Carty57,Garde69} suggest that the air resistance coefficient for a sphere rolling on a plane is larger than its usual value $C_D=0.45$ in unbounded fluid. For simplicity, we used $C_D=1$ in the following. Since it is hard to know precisely the value of the  kinetic friction coefficient in our experiment, $\mu$ is  kept as a fitting parameter but serves only to scale the value of our experimental parameter $N$. In the following, $\mu=0.02$ is used, which is slightly smaller than the value expected for a lubricated metal-plastic interface.

Figure \ref{rescaling} shows our experimental data for mild steel beads compared to the
model described above. In this figure, bead velocities have been rescaled by the backward solution $V_0$ (eq.\ref{solutionA}) and plotted as a function of our dimensionless number $N$. Although it involves a variety of values for $h$ and $a$, our experimental data show a very good
agreement with both  solutions (\ref{solutionA}) and (\ref{solutionB}),
represented by the dashed lines. The bifurcation from the rolling
solution to the sliding state occurring at $N=1$ is relatively sharp, and very weak departure from theory is observed, although $4$ orders of magnitude for $N$ are explored.\\

\begin{figure}
\centerline{\includegraphics[width= 9 cm]{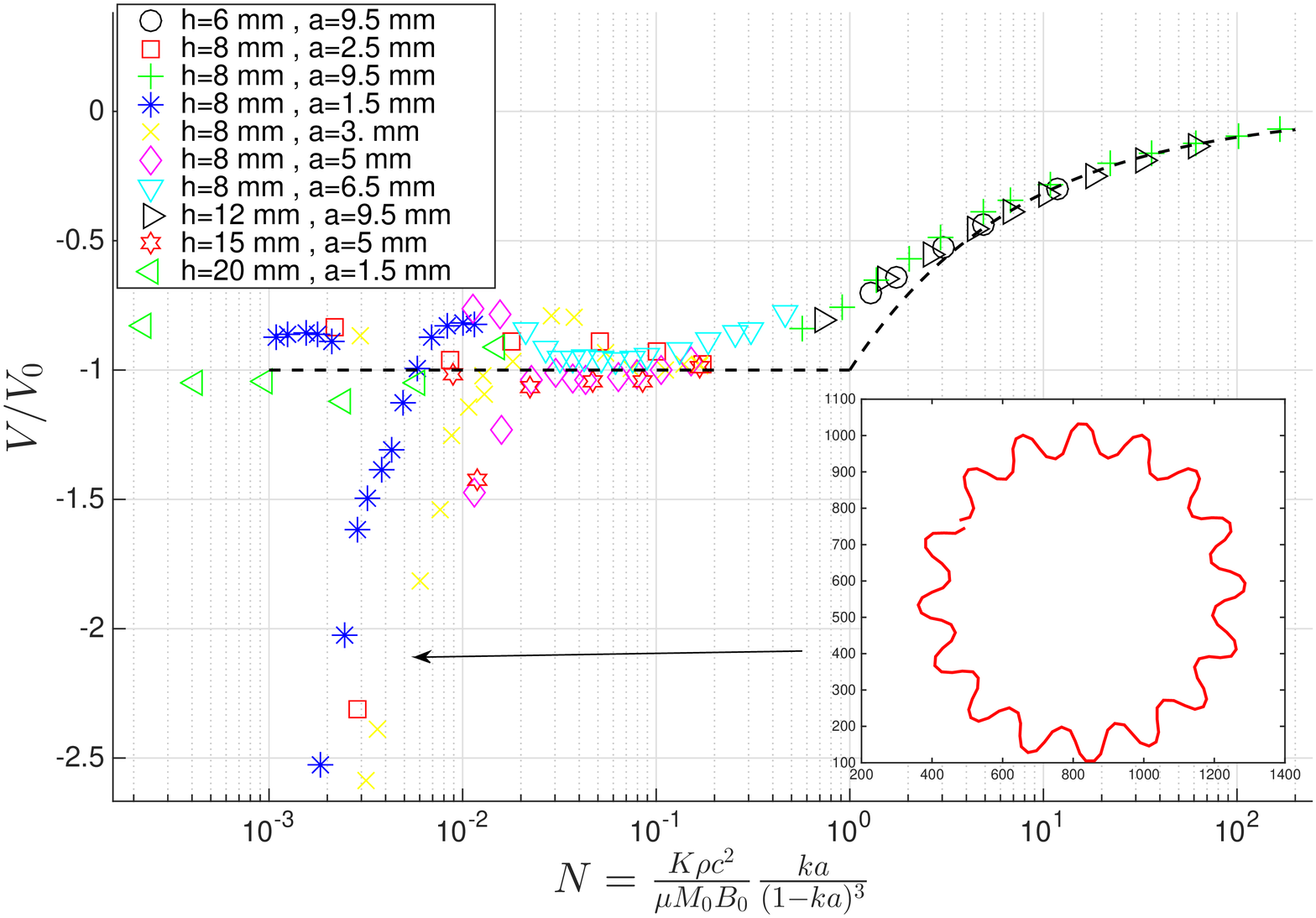} }
\caption{Evolution of $V/V_0$ as a function of $N$ for NdFeB spheres. At small $N$, a bistability is observed between the rolling backward solution $V_0$  and a faster motion. Inset: corresponding bidimensionnal oscillating trajectories.}
\label{rescaling2}
\end{figure}

For Neodymium magnets (Fig.\ref{rescaling2}), the rescaling also shows a fairly good agreement with the theory at large $N$, but exhibits clear departures from the model at small $N$.
 First, the transition from rolling to sliding is strongly imperfect, involving a smoother transition and  smaller velocities than expected. Moreover, the data are 
spread around the backward-rolling solution. In particular, note that
some experiments involve backward velocities several times faster than $V_0$ as $N$ is reduced. The inset of Fig.\ref{rescaling2}, which focus on the
trajectory obtained for $h=8$mm and $a=3$mm, shows that
this departure is due to a more complex trajectory than the
one-dimensional motion used during the model derivation: the bead
starts oscillating radially around its circular trajectory. It is interesting
to note that this type of motion which involves a complex interaction
between the magnetic restoring force  and the inertia of
the sphere, can increase the velocity of the bead even further . Clearly, this motion would require a modification of our 1-D model, which is beyond the scope of this paper. 

Additional experiments were done using beads made of copper: in this non-magnetic case, a similar counter-motion is observed, although one order of magnitude smaller than the one reported here for magnetic spheres. In this case, eddy currents in the sphere produce weaker but similar forces than the magnetic ones described here, thus leading to identical behavior. Finally, note that by adding random noise in equations (\ref{eq1})-(\ref{eq2}), these equations easily explain the origin of the chaotic  behavior reported in Fig.\ref{hysteresis}: it originates from imperfections of the bottom surface, generating noise in the friction and allowing random transitions between the two bistable forward/backward solutions. \\

\section{Conclusion}

The motion of a magnetized sphere driven by an external travelling magnetic field has been studied. When the speed of the driving wave is large enough, we observe a transition from a state involving a forward synchronous translation with no rotation, to a backward translation associated to synchronous angular rotation. Under several assumptions, a simple model was proposed to explain this backward motion, which can be several times faster than the driving wave. A good agreement is obtained between the theory and experimental data.

The behavior reported in this paper may have several applications. First, the present experiment shares some similarities with recent studies on the control of magnetic beads in microfluidic channel \cite{Karle2011},\cite{Mahonet2011} and  may therefore be regarded as a new method for manipulating nano-microscale objects using the tumbling motion of particles induced by a travelling magnetic wave. Rolling of a small magnetic particle can also be relevant  to micromanipulation of magnetic beads for magnetic twizzer \cite{Yan2004} or polishing techniques.  It is also interesting that the present mechanism offers a simple  explanation to counter-flows  observed near boundaries  in ferrofluids submitted to a rotating magnetic field. On the other hand, the mechanism of locomotion described here is very general, and may be applied to controlled transport of objects at macroscopic scales. In this perspective, several questions may be addressed in future work, such as the role of electrical conductivity  or the factors limiting the stability of the counter-propagative solution.

%
%




\acknowledgments
The author is thankful to S. Fauve, F. P\'etr\'elis, G. Michel and B. Semin for their comments and criticisms,
which helped to improve the manuscript. This work was supported by funding from the French program "Retour Postdoc" managed by Agence Nationale de la Recherche (Grant ANR-398031/1B1INP)

\end{document}